%% file: ANA-HIGG-2021-23-PAPER.tex
\documentclass[PAPER, atlasdraft=false, texlive=2023, UKenglish, cernpreprint, texmf, orcidlogo]{atlasdoc}
 
\usepackage{atlaspackage}
\usepackage{atlasbiblatex}
 
\usepackage{atlasphysics}

\usepackage{xspace}
\usepackage{multirow}

\graphicspath{{logos/}{figures/}}
 
\usepackage{ANA-HIGG-2021-23-PAPER-defs}

\AtlasTitle{A detailed map of Higgs boson interactions by the ATLAS experiment ten years
after the discovery}

\AtlasAbstract{
The Standard Model of particle physics~\cite{Weinberg:1967tq, Glashow:1961tr,
Salam:1968rm, tHooft:1972fi} describes the known fundamental particles and  forces that
make up our universe, with the exception of gravity.  One of the central features of the
Standard Model is a
field that permeates all of space and interacts with fundamental
particles~\cite{Higgs:1964ia, Higgs:1964pj, Englert:1964et, Guralnik:1964eu,
Kibble:1967sv}. The quantum excitation of this field, known as Higgs field, manifests
itself as the Higgs boson, the only fundamental particle with no spin.  In 2012, a
particle with properties consistent with the Higgs boson of the Standard Model was
observed by the ATLAS and CMS experiments at the Large Hadron Collider at
CERN~\cite{HIGG-2012-27,CMS-HIG-12-028}.   Since then, more than 30 times as many Higgs
bosons have been recorded by the ATLAS experiment, allowing much more precise measurements
and new tests of the theory. Here, on the basis of this larger dataset, we combine an
unprecedented number of production and decay processes of the Higgs boson to scrutinize
its interactions with elementary particles. Interactions with gluons, photons, and $W$ and
$Z$ bosons -- the carriers of the strong, electromagnetic, and weak forces -- are studied in detail.
Interactions with three third-generation matter particles (bottom ($b$) and top ($t$)
quarks, and tau leptons ($\tau$)) are well
measured and indications of interactions with a second-generation particle (muons, $\mu$) are
emerging. These tests reveal that the Higgs boson discovered ten years ago is remarkably
consistent with the predictions of the theory and provide stringent constraints on many
models of new phenomena beyond the Standard Model. }

\AtlasRefCode{HIGG-2021-23}
 
\PreprintIdNumber{CERN-EP-2022-057}

\AtlasJournal{Nature}
\AtlasJournalRef{Nature 607, 52–59 (2022)}
\AtlasDOI{10.1038/s41586-022-04893-w}

\hypersetup{pdftitle={A detailed map of Higgs boson interactions by the ATLAS experiment ten years
after the discovery},pdfauthor={The ATLAS Collaboration}}
 
\begin{document}
 
\maketitle

\newpage

% The next lines are included from the .//sections/00_summary.tex input file

% End of text imported from the .//sections/00_summary.tex input file

% The next lines are included from the .//sections/01_introduction.tex input file
The Standard Model (SM) of particle physics has been tested by many experiments since its
formulation~\cite{Weinberg:1967tq, Glashow:1961tr, Salam:1968rm, tHooft:1972fi} and, after
accounting for the neutrino masses, no discrepancies between experimental observations and
its predictions have been established so far. A central feature of the SM is the existence
of a spinless quantum field that permeates the universe and gives mass to massive
elementary particles. Testing the existence and properties of this field and its
associated particle, the Higgs boson, has been one of the main goals of particle physics
for several decades. In the SM, the strength of the interaction, or `coupling', between
the Higgs boson and a given particle is fully defined by the particle's mass and type.
There is no direct coupling to the massless SM force mediators, the photons and gluons,
whereas there are three types of couplings to massive particles in the theory. The first is
the `gauge' coupling of the Higgs boson to the mediators of the weak force, the $W$ and
$Z$ vector bosons. Demonstrating the existence of gauge couplings is an essential test of
the
spontaneous electroweak symmetry breaking mechanism~\cite{Higgs:1964ia,
Higgs:1964pj, Englert:1964et, Guralnik:1964eu, Kibble:1967sv}. The second type of coupling involves another kind of
fundamental interaction, the Yukawa interaction, between the Higgs boson and matter
particles, or fermions.  The third type of coupling is the `self-coupling' of the Higgs
boson to itself. A central prediction of the theory is that the couplings scale with the
particle masses, and they are all precisely predicted once all the particle masses are
known. The experimental determination of the couplings of the Higgs boson to each individual particle therefore provides important and independent tests of the SM. It also provides stringent constraints on theories beyond the SM, which generally predict different patterns of coupling values.
 
In 2012, the ATLAS~\cite{PERF-2007-01} and CMS~\cite{CMS-CMS-00-001} experiments at the
Large Hadron Collider (LHC)~\cite{Evans:2008zzb} at CERN announced the discovery of a new
particle with properties consistent with those predicted for the Higgs boson of the
SM~\cite{HIGG-2012-27,CMS-HIG-12-028}. More precise measurements that used all of the
proton--proton collision data taken during the first data-taking period from 2011 to 2012
at the LHC (Run~1) showed evidence that, in contrast to all other known fundamental
particles, the properties of the discovered particle were consistent with the hypothesis that it
has no spin~\cite{HIGG-2013-17,CMS-HIG-14-018}.  Alternate spin 1 and spin 2 hypotheses were also tested and were
excluded at a high level of confidence.
Investigations of the charge conjugation and parity properties of the new particle were
also performed, demonstrating consistency with the CP-even quantum state predicted by the
SM, while still allowing for small admixtures of non-SM CP-even or CP-odd states~\cite{HIGG-2013-17, CMS-HIG-14-018}. Limits on its lifetime were obtained through indirect measurements of
its natural width~\cite{HIGG-2013-17,CMS-HIG-14-018,HIGG-2014-10, CMS-HIG-14-032, CMS-HIG-14-036}. In addition, more precise measurements of the new particle’s interactions with other elementary particles were achieved~\cite{HIGG-2015-07}.  The results of all these investigations demonstrated that its properties were compatible with those of the SM Higgs boson.
However, the statistical uncertainties associated with these early measurements allowed
considerable room for possible interpretations of the data in terms of new phenomena beyond the SM and left many predictions of the SM untested.
 
The characterization of the Higgs boson continued during the Run~2 data-taking period between 2015 and 2018.  About 9~million Higgs bosons are predicted to have been produced in the ATLAS detector during this period, of which only about 0.3\% are experimentally accessible. This is 30 times more events than at the time of its discovery, due to the higher rate of collisions and the increase of the collision energy from 8~teraelectronvolts~(\TeV) to 13~\TeV, which raises the production rate.
In this Article, the full Run 2 dataset, corresponding to an integrated luminosity of
139 inverse femtobarns (\ifb), is used for the measurements of
Higgs boson production and decay rates, which are used to study the couplings between the Higgs boson and the particles involved.
This improves on the previous measurements obtained with partial Run 2 datasets~\cite{CMS-HIG-17-031, HIGG-2018-57}.
The corresponding predictions depend on the value of the Higgs boson mass, which has now
been measured by the ATLAS and CMS experiments~\cite{HIGG-2014-14, HIGG-2016-33,
CMS-HIG-19-004} with an uncertainty of approximately 0.1\%. The predictions employed in this Article use the combined central value of 125.09~\GeV~\cite{HIGG-2014-14}.
 
The dominant production process at the LHC, which accounts for about 87\% of Higgs boson
production, is the heavy-quark loop-mediated gluon--gluon fusion process (ggF). The second most copious process is vector-boson fusion (VBF) where two weak bosons, either $Z$ or $W$ bosons, fuse to produce a Higgs boson (7\%).
Next in rate is production of a Higgs boson in association with a weak ($V = W, Z$) boson (4\%).
Production of a Higgs boson in association with a pair of top quarks ($t\bar{t}H$) or
$b$-quarks ($b\bar{b}H$) each account for about 1\% of the total rate. The contribution of
other $q\bar{q}H$ processes is much smaller and experimentally not accessible. Only about 0.05\%
of Higgs bosons are produced in association with a single top quark ($tH$). Representative
Feynman diagrams of these processes are shown in Figures~\ref{fig:feynman_graphs}(a)--\ref{fig:feynman_graphs}(e).
After it is produced, the Higgs boson is predicted to decay almost instantly, with a
lifetime of $1.6\times10^{-22}$ seconds. More than 90\% of these decays are via eight
decay modes (see Figures~\ref{fig:feynman_graphs}(f)--\ref{fig:feynman_graphs}(i)): decays
into gauge boson pairs, i.e.\  $W$ bosons with a probability, or branching fraction, of
22\%,  $Z$ bosons 3\%,  photons ($\gamma$) 0.2\%,  $Z$ boson and photon 0.2\%, as well as
decays into fermion pairs, i.e.\  $b$-quarks 58\%,   $c$-quarks 3\%, $\tau$-leptons 6\%,
and muons ($\mu$) 0.02\%. There may also be decays of the Higgs boson into invisible
particles, above the SM prediction of 0.1\%,  which are also searched for. Such decays are
possible in theories beyond the SM, postulating, for example, the existence of dark-matter particles which do not interact with the detector.

\begin{figure}[!htbp]
\centering
\includegraphics[width=1.0\textwidth]{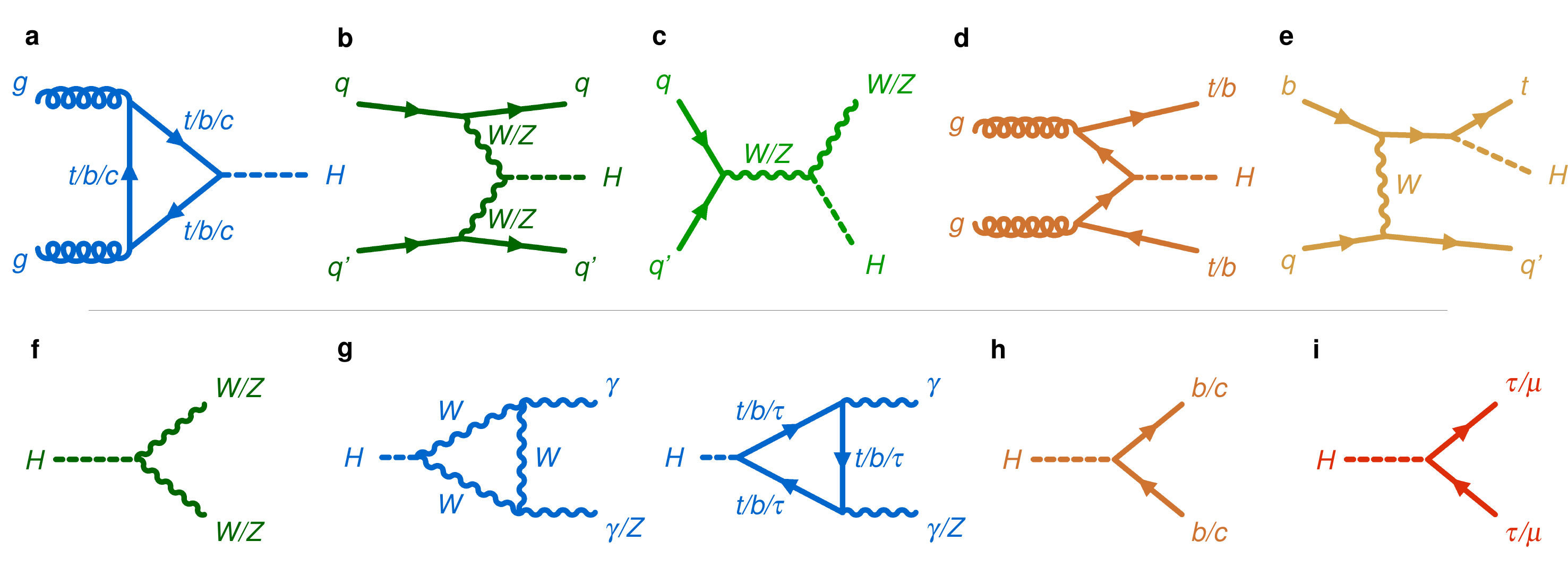}
\caption{
\textbf{Examples of Feynman diagrams for Higgs boson  production
and decays.} The Higgs boson is produced via gluon--gluon fusion (a),
vector-boson fusion (VBF; b),  and associated production with vector bosons (c), top- or
$b$-quark pairs (d), or a single top quark (e). f--i, The Higgs boson decays into a pair of
vector bosons (f), a pair of photons or a $Z$ boson and a photon (g), a pair of quarks (h),
and a pair of charged leptons (i). Loop-induced Higgs boson interactions with gluons or
photons are shown in blue, processes involving couplings to $W$ or $Z$ bosons in green, to
quarks in orange, and to leptons in red. Two different shades of green (orange) are used
to separate the VBF and $VH$ ($\ttbar H$ and $tH$) production processes.}
\label{fig:feynman_graphs}
\end{figure}
 
In this Article, the mutually exclusive measurements of Higgs boson production and decays probing all processes listed above are combined, taking into account the correlations among their uncertainties. In a single measurement, different couplings generally contribute in the production and decay. The combination of all measurements is therefore necessary to constrain these couplings individually. This allows key tests of the Higgs sector of the SM to be performed, including the determination of the coupling strengths of the Higgs boson to various fundamental particles and a comprehensive study of the kinematic properties of Higgs boson production. The latter could reveal new phenomena beyond the SM that are not observable through measurements of the coupling strengths.

% End of text imported from the .//sections/01_introduction.tex input file

\vspace{-15pt}
\section*{The ATLAS detector at the LHC}
\vspace{-5pt}

% The next lines are included from the .//sections/02_ATLAS.tex input file
The ATLAS experiment~\cite{PERF-2007-01} at the LHC is a multipurpose particle detector with a forward--backward symmetric, cylindrical geometry and a near \(4\pi\) coverage in solid angle. The detector records digitized signals produced by the products of LHC's proton bunch collisions, hereafter termed collision `events'. It is designed to identify a wide variety of particles and measure their momenta and energies. These particles include electrons, muons and $\tau$-leptons, photons, as well as gluons and quarks which produce collimated jets of particles in the detector. Since the jets from $b$-quarks and $c$-quarks contain hadrons with relatively long lifetimes, they can be identified by observing a decay vertex which typically occurs at a measurable distance from the collision point. The presence of particles that do not interact with the detector, such as neutrinos, can be inferred by summing the vector momenta of the visible particles in the plane transverse to the beam and imposing conservation of transverse momenta.
 
The detector components  closest to the collision point measure  charged-particle
trajectories and momenta. This inner spectrometer is surrounded by calorimeters that are
used in the identification of particles and in the measurement of their energies. The
calorimeters are in turn surrounded by an outer spectrometer dedicated to measuring the trajectories and
momenta of muons, the only charged particle to travel through the calorimeters. A two-level
trigger system was optimized for Run~2 data-taking~\cite{TRIG-2019-04} to select events of
interest at a rate of about \SI{1}{\kHz} from the proton bunch collisions occurring at a
rate of 40 MHz.
An extensive software suite~\cite{ATL-SOFT-PUB-2021-001} is used in the simulation, reconstruction and analysis of real and simulated data, in detector operations, and in the trigger and data acquisition systems of the experiment.

% End of text imported from the .//sections/02_ATLAS.tex input file

\vspace{-15pt}
\section*{Input measurements and combination procedure}
\vspace{-5pt}

% The next lines are included from the .//sections/03_00_channels.tex input file
Physics analyses typically focus on particular production and decay processes and measure the number of Higgs boson candidates observed after accounting for non-Higgs background processes. To determine the strengths of the interactions of the Higgs boson, simultaneous fits with different physically motivated assumptions are performed on a combined set of complementary measurements. The relative weights of the input measurements in the combination depend on the analysis selection efficiencies, on the signal rates associated with the Higgs processes studied by the analysis, on the signal-to-background ratios, and on the associated systematic uncertainties.
 
For each decay mode entering the combination, the production process is assessed via event classification based upon the properties of particles produced in association with the Higgs boson, mostly via dedicated machine-learning approaches. Unless stated otherwise, studies of each decay mode consider all individual or combined contributions from six production processes: ggF, VBF, $WH$, $ZH$, $t\bar{t}H$ and $tH$. Higgs boson interactions are further explored via additional event classification of each production process based on the kinematic properties of the produced Higgs boson and the associated particles.
 
The input to the combined measurement includes the latest results from the decay modes
which initially led to the Higgs boson discovery:  $H\to ZZ\to \ell^+\ell^-\ell^+\ell^-$
decays~\cite{HIGG-2018-28} with two $Z$ bosons that subsequently decay into a pair of
oppositely charged electrons or muons; $H\to W^{\pm}W^{\mp}\to
\ell^\pm\nu_\ell\ell^\mp\nu_\ell$ decays targeting separately the ggF and
VBF~\cite{HIGG-2021-20-bib}, and $WH$ and $ZH$~\cite{HIGG-2017-14} production processes;
and $H\to \gamma\gamma$ decays~\cite{HIGG-2020-16} with two high-energy photons. The latter is the only measurement used to discriminate between the $t\bar{t}H$ and $tH$ processes.
These diboson decay modes are for the first time complemented by a search for the rare $H\to Z\gamma \to \ell^+\ell^-\gamma$ decay~\cite{HIGG-2018-42}.
The decays of Higgs bosons to fermions are also extensively explored. The measurement of the dominant $H\to b\bar{b}$ decay mode is particularly challenging due to a very large multi-jet background, which can be suppressed by requiring the presence of additional particles characteristic of the $WH$ or $ZH$~\cite{HIGG-2018-51, HIGG-2018-52}, VBF~\cite{HIGG-2019-04} and $t\bar{t}H$~\cite{HIGG-2020-23} production processes. As a new input, the fully hadronic $H\to b\bar{b}$  signal events with large Higgs boson transverse momentum are also considered~\cite{HIGG-2021-08}, providing for the first time sensitivity to the ggF production process in this decay mode.
The sensitivity of the latest measurement in the $H\to \tau^+\tau^-$ decay
mode~\cite{HIGG-2019-09} is now extended to the $VH$ and combined $t\bar{t}H$ and $tH$
production processes.  In addition to the $t\bar{t}H$ measurements obtained in the $\gamma\gamma$, $\tau^+\tau^-$ and $ZZ$ decay modes, a complementary analysis which is sensitive to  $\tau^+\tau^-$, $W^{\pm}W^{\mp}$ and $ZZ$ decays is performed using events with multiple leptons in the final state~\cite{HIGG-2017-02}.
The considerably more challenging measurements of Higgs boson couplings to second-generation fermions are explored via searches for the $H\to \mu^+\mu^-$~\cite{HIGG-2019-14} and, included in the combination for the first time, $H\to c\bar{c}$~\cite{HIGG-2021-12} decays. Due to the large multi-jet background, the latter decay mode is currently accessed only via $WH$ and $ZH$ production.
Finally, the inputs to the combination are complemented by the latest direct searches in
the VBF and $ZH$ production processes for Higgs boson decays into invisible particles
which escape the detector~\cite{EXOT-2020-11, HIGG-2018-26}. A summary of these
input measurements used in the combination is available in Extended Data Table 1.

All input measurements are performed with the full set of Run~2 data, except for the measurements in Refs.~\cite{HIGG-2017-14, HIGG-2017-02}, which use a partial Run~2 dataset collected during 2015 and 2016. The direct searches for invisible Higgs boson decays and the $H\to c\bar{c}$ measurements are only employed for measurements of the relevant Higgs boson coupling strengths, while the $H\to b\bar{b}$ measurements at high Higgs boson transverse momenta~\cite{HIGG-2021-08} are only considered when probing the kinematic properties of Higgs boson production. All other inputs are used for the measurements of production cross sections, branching fractions and coupling strengths. The measurement of kinematic properties of Higgs boson production excludes input measurements from Refs.~\cite{HIGG-2017-14, HIGG-2017-02, HIGG-2018-42, HIGG-2019-14, HIGG-2021-12} due to their limited sensitivity.
 
Analyses performed with the Run~2 data introduce a number of improvements, often resulting in up to 50\% better signal sensitivities compared to those expected from just the increase in the analysed amount of data. These improvements include better particle reconstruction that was optimized to cope with an increased number of proton interactions per bunch crossing, dedicated reconstruction of highly Lorentz boosted $H\to b\bar{b}$ decays, a greater number of simulated events, higher granularity of the kinematic regions that are probed in each production process, and improved signal and background theory predictions.
 
The SM is tested by comparing the observed signal rates to theory predictions that require state-of-the-art calculations of Higgs boson production cross sections and branching fractions (see Refs.~\cite{YR4, Lidnert:2018kmw, Djouadi:2018kms, Bonetti:2018ukf, Dulat:2018rbf, Harlander:2018yio, Cacciari:2018dksz}). All signal reconstruction efficiencies and most background rates are predicted from the simulation. The simulation is complemented by the use of dedicated signal-depleted control data for measurements of selected background processes and to constrain signal selection efficiencies. A common set of event generators were used in all analyses to describe the gluon and quark interactions in the proton--proton collisions. The generated particles were passed through a detailed simulation of the ATLAS detector response prior to their reconstruction and identification.

% End of text imported from the .//sections/03_00_channels.tex input file

% The next lines are included from the .//sections/04_stats.tex input file
The statistical analysis of the data is described in more detail in Methods. It relies on a likelihood formalism, where the
product of the likelihood functions describing each of the input measurements is calculated in order to obtain a
combined likelihood~\cite{ATL-PHYS-PUB-2011-011}. The effects of experimental and theoretical systematic uncertainties on the predicted
signal and background yields are implemented by including nuisance parameters in the
likelihood function. The values of those additional parameters are either fully determined
by the included data, or constrained by Gaussian terms that multiply the
likelihood. The effects of uncertainties that affect multiple measurements are propagated coherently through the
fit by using common nuisance parameters.
 
The statistical test of a given signal hypothesis, used for the measurement of the
parameters of interest, is performed with a test statistic based on the profile likelihood
ratio~\cite{Cowan:2010st}. The confidence intervals of the measured parameters and the $p$-value used to test the compatibility of the results and the SM predictions are constructed from the test statistic distribution, which is obtained using asymptotic formulae~\cite{Cowan:2010st}.
 
The total uncertainty in the measurement of a given parameter of interest can be decomposed into different components.
The statistical uncertainty is obtained from a fit with all externally constrained nuisance parameters set to their best-fit values.
The systematic uncertainty, whose squared value is evaluated as the difference between the squares of the total uncertainty and the statistical uncertainty, can be decomposed into categories by setting all relevant subsets of nuisance parameters to their best-fit values.

% End of text imported from the .//sections/04_stats.tex input file

\vspace{-15pt}
\section*{Combined measurement with ATLAS Run 2 data}
\vspace{-5pt}

% The next lines are included from the .//sections/05_results.tex input file

% The next lines are included from the .//sections/05_results_mu.tex input file
The Higgs boson production rates are probed by the likelihood fit to observed signal yields described earlier. As the production cross section $\sigma_i$ and the branching fraction $B_f$ for a specific production process $i$ and decay mode $f$ cannot be measured separately without further assumptions, the observed signal yield for a given process is expressed in terms of a single signal-strength modifier $\mu_{if} = \left(
\sigma_i / \sigma_i^\text{SM} \right) \times \left( \BR_f / \BR_f^\text{SM} \right)$,
where the superscript `SM' denotes the corresponding SM prediction.
Assuming that all production and decay processes scale with the same global signal strength $\mu = \mu_{if}$, the inclusive Higgs boson production rate relative to the SM prediction is measured to be
\begin{equation*}
\mu =1.05 \pm 0.06 = 1.05\pm 0.03\, (\text{stat.})\, \pm 0.03\, (\text{exp.})\, \pm 0.04\, (\text{sig.\ th.})\, \pm 0.02\, (\text{bkg.\ th.}).
\end{equation*}
The total measurement uncertainty is decomposed into components for statistical
uncertainties, experimental systematic uncertainties, and theory uncertainties in both
signal and background modelling.  Both the experimental and the theoretical uncertainties
are almost a factor of two lower than in the Run~1 result~\cite{HIGG-2015-07}. The
presented measurement supersedes the previous ATLAS combination with a partial Run~2
dataset~\cite{HIGG-2018-57}, decreasing the latest total measurement uncertainty by about
30\%.

% End of text imported from the .//sections/05_results_mu.tex input file

% The next lines are included from the .//sections/05_results_xsbr.tex input file
Higgs boson production is also studied per individual process.
As opposed to the top-quark decay products from $\ttbar H$ production,  the identification
efficiency of $b$-jets from the $\bbbar H$ production is low, making the $\bbbar H$ process
experimentally indistinguishable from ggF production. The $\bbbar H$ and ggF processes are
therefore grouped together, with $\bbbar H$ contributing by a relatively small amount of the
order of 1\% to the total ggF$+\bbbar H$ production.
In cases where several processes are combined, the combination assumes the
relative fractions of the components to be those from the SM within corresponding theory
uncertainties. Results are obtained from the fit to the data, where the cross section of
each production process is a free parameter of the fit. Higgs boson decay branching
fractions are set to their SM values, within the uncertainties specified in
Ref.~\cite{YR4}. The results are shown in Figure~\ref{fig:sixxs_br_obs}(a).

\begin{figure}[!htbp]
\centering
\subfloat[]{\includegraphics[width=.5\textwidth]{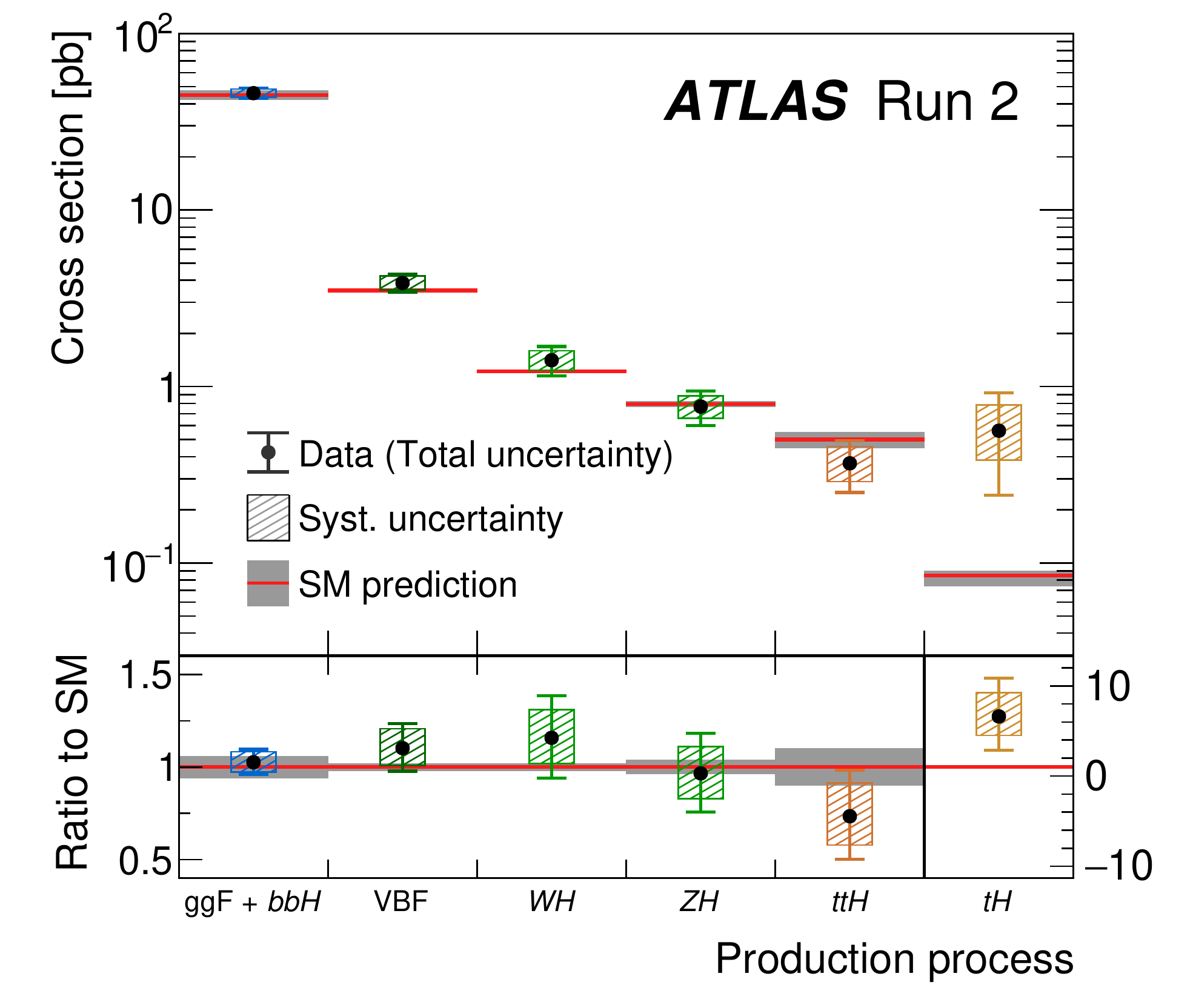}}
\subfloat[]{\includegraphics[width=.5\textwidth]{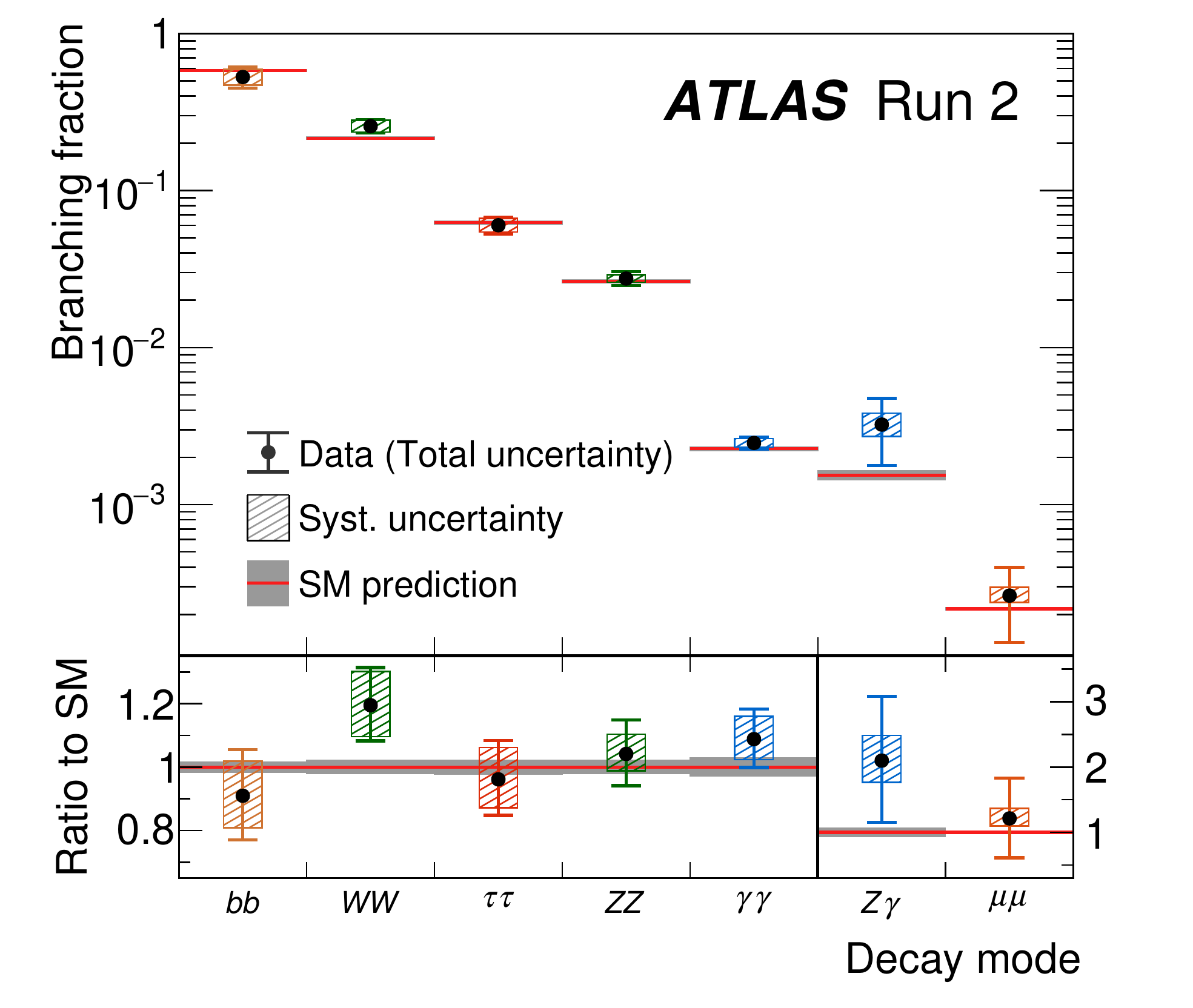}}
\caption{\textbf{Observed and predicted Higgs boson production cross-sections and
branching fractions.} a, Cross sections for different Higgs boson production
processes are measured assuming SM values for the decay branching fractions.
b, Branching fractions for different Higgs boson decay modes are measured assuming SM values for the production cross sections.  The lower panels show the ratios of the measured values to their SM predictions. The vertical bar on each point denotes the 68\% confidence interval. The $p$-value for compatibility of the measurement and the SM prediction is 65\% for a and 56\% for b.}
\label{fig:sixxs_br_obs}
\end{figure}
 
All measurement results are compatible with the SM predictions. For the ggF and VBF
production processes, which were already observed in Run~1 data, the cross sections are
measured with a precision of 7\% and 12\%, respectively. The following production
processes are now also observed:  $WH$ with an observed (expected) signal significance
of 5.8 (5.1) standard deviations ($\sigma$), $ZH$ with 5.0$\sigma$
(5.5$\sigma$) and the combined $t\bar{t}H$ and $tH$ production processes with 6.4$\sigma$
(6.6$\sigma$), where the expected signal significances are obtained under the SM hypothesis. The separate $t\bar{t}H$ and
$tH$ measurements lead to an observed (expected) upper limit on $tH$ production of 15~(7) times the SM prediction at the 95\% confidence level (CL), with a relatively large negative
correlation coefficient of 56\% between the two measurements. This is due to
cross-contamination between the $t\bar{t}H$ and $tH$ processes in the set of reconstructed
events that provide the highest sensitivity to these production processes.

Branching fractions of individual Higgs boson decay modes are measured by setting the cross sections
for Higgs boson production processes to their respective SM values. The results are shown in Figure~\ref{fig:sixxs_br_obs}(b).
The branching fractions of the $\gamma\gamma$, $ZZ$, $W^{\pm} W^{\mp }$ and
$\tau^+\tau^-$ decays, which were already observed in the Run~1 data, are measured with
a precision ranging from 10\% to 12\%. The $b\bar{b}$ decay mode is
observed with a signal significance of 7.0$\sigma$ (expected 7.7$\sigma$), while the observed (expected) signal significances for the $H\to \mu^+\mu^-$ and $H\to Z\gamma$ decays are 2.0$\sigma$ (1.7$\sigma$) and 2.3$\sigma$ (1.1$\sigma$), respectively.

The assumptions about the relative contributions of different decay or production processes in the above measurements are relaxed by directly measuring the
product of production cross section and branching fraction for different combinations of
production and decay processes. The corresponding results are shown in Figure~\ref{fig:xsbr_obs}.
The measurements are in agreement with the SM prediction.
\begin{figure}[!htbp]
\centering
\includegraphics[width=\textwidth]{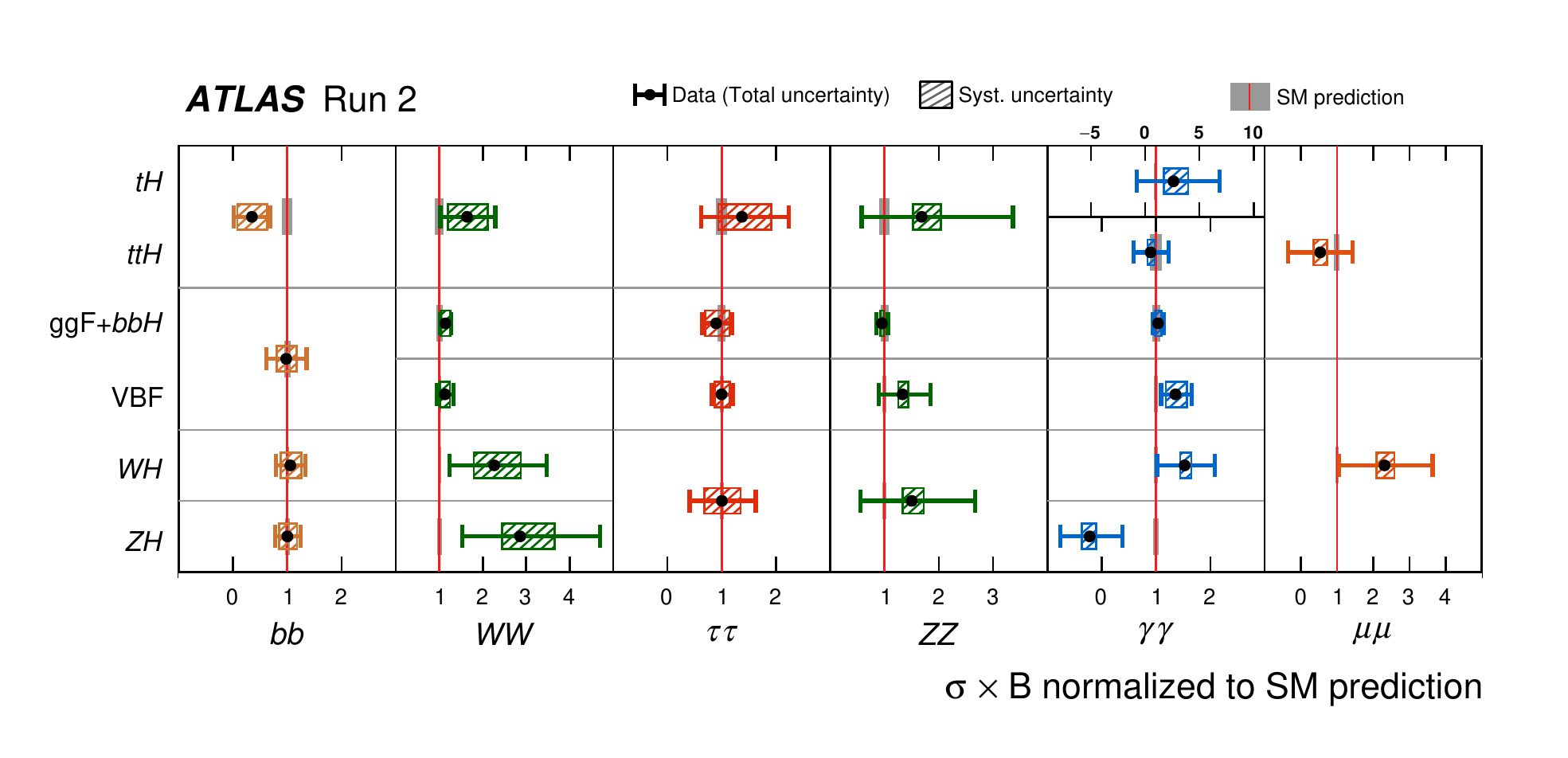}
\caption{\textbf{Ratio of observed rate to predicted SM event rate for different combinations of
Higgs boson production and decay processes.}  The horizontal bar on each point denotes the 68\% confidence interval. The narrow grey
bands indicate the theory uncertainties in the SM cross-section times the branching
fraction predictions.
The $p$-value for compatibility of the measurement and the SM prediction is
72\%.}
\label{fig:xsbr_obs}
\end{figure}

% End of text imported from the .//sections/05_results_xsbr.tex input file

% The next lines are included from the .//sections/05_results_kappa.tex input file
To determine the value of a particular Higgs boson coupling strength, a simultaneous fit of many individual production times branching fraction measurements is required. The coupling fit presented here is performed within the $\kappa$-framework~\cite{Heinemeyer:2013tqa} with a set of parameters $\vec{\kappa}$ that affect the Higgs boson coupling strengths without altering any kinematic distributions of a given process.
 
Within this framework, the cross section times the branching fraction for an individual measurement is parameterized in terms of the multiplicative coupling strength modifiers $\kappa$. A coupling strength modifier $\kappa_p$ for a production or decay process via the coupling to a given particle $p$ is defined as $\kappa_p^2 = \sigma_p / \sigma_p^\text{SM}$ or $\kappa_p^2 = \Gamma_p/\Gamma_p^\text{SM}$, respectively, where $\Gamma_p$ is the partial decay width into a pair of particles $p$. The parameterization takes into account that the total decay width depends on all decay modes included in the present measurements, as well as currently undetected or invisible, direct or indirect decays predicted by the SM (such as those to gluons, light
quarks, or neutrinos) and the hypothetical decays into non-SM particles. The decays to non-SM particles are divided into decays to invisible particles and other decays that would be undetected due to large backgrounds. The corresponding branching fractions for the two are denoted by
$B_{\mathrm{inv.}}$ and $B_{\text{u.}}$, respectively.

In the following, three classes of models with progressively fewer assumptions about coupling strength modifiers are considered. SM values are assumed for the coupling strength modifiers of first-generation fermions, and the modifiers of the second-generation quarks are set to those of the third generation except where $\kappa_c$ is left free-floating in the fit. Due to their small sizes, these couplings are not expected to noticeably affect any of the results. The ggF production and the $H \to \gamma\gamma$ and $H \to Z\gamma$ decays are loop-induced processes. They are either expressed in terms of the more fundamental coupling strength scale factors corresponding to the particles that contribute to the loop-induced processes in the SM, or treated using effective coupling strength modifiers $\kappa_g$, $\kappa_\gamma$ and $\kappa_{Z\gamma}$, respectively. The latter scenario accounts for possible loop contributions from particles beyond the SM. The small contribution from the loop-induced $gg\to ZH$ process is always parameterized in terms of the couplings to the corresponding SM particles.

The first model tests one scale factor for the vector bosons, $\kappa_V=\kappa_W =
\kappa_Z$, and a second, $\kappa_F$, which applies to all fermions.
In general, the SM prediction of $\kappa_V=\kappa_F = 1$ does not hold in extensions of the SM. For example, the values of $\kappa_V$ and $\kappa_F$ would be less than 1 in models in which the Higgs boson is a composite particle.
The effective couplings
corresponding to the ggF, $H \to \gamma\gamma$ and $H \to Z\gamma$ loop-induced processes are
parameterized in terms of the fundamental SM couplings. It is assumed that there are no
invisible or undetected Higgs boson decays beyond the SM, i.e.~$B_{\mathrm{inv.}} =
B_{\text{u.}} = 0$. As only the relative sign between $\kappa_V$ and $\kappa_F$ is
physical and a negative relative sign has been excluded with a high level of
confidence~\cite{HIGG-2015-07}, $\kappa_V \ge 0$ and $\kappa_F \ge 0$ are assumed.
Figure~\ref{fig:kf_kv} shows the results of a combined fit in the
($\kappa_V$,~$\kappa_F$)~plane. The best-fit values and their uncertainties from the
combined fit are $\kappa_V = 1.035 \pm 0.031$ and  $\kappa_F = 0.95 \pm 0.05$, compatible
with the SM predictions. A relatively large positive correlation of 39\% is observed between the two fit parameters, since some of the most sensitive input measurements involve the ggF production process (i.e. via couplings to fermions) with subsequent Higgs boson decays into vector bosons.
\begin{figure}[h!]
\centering
\includegraphics[width=.45\textheight]{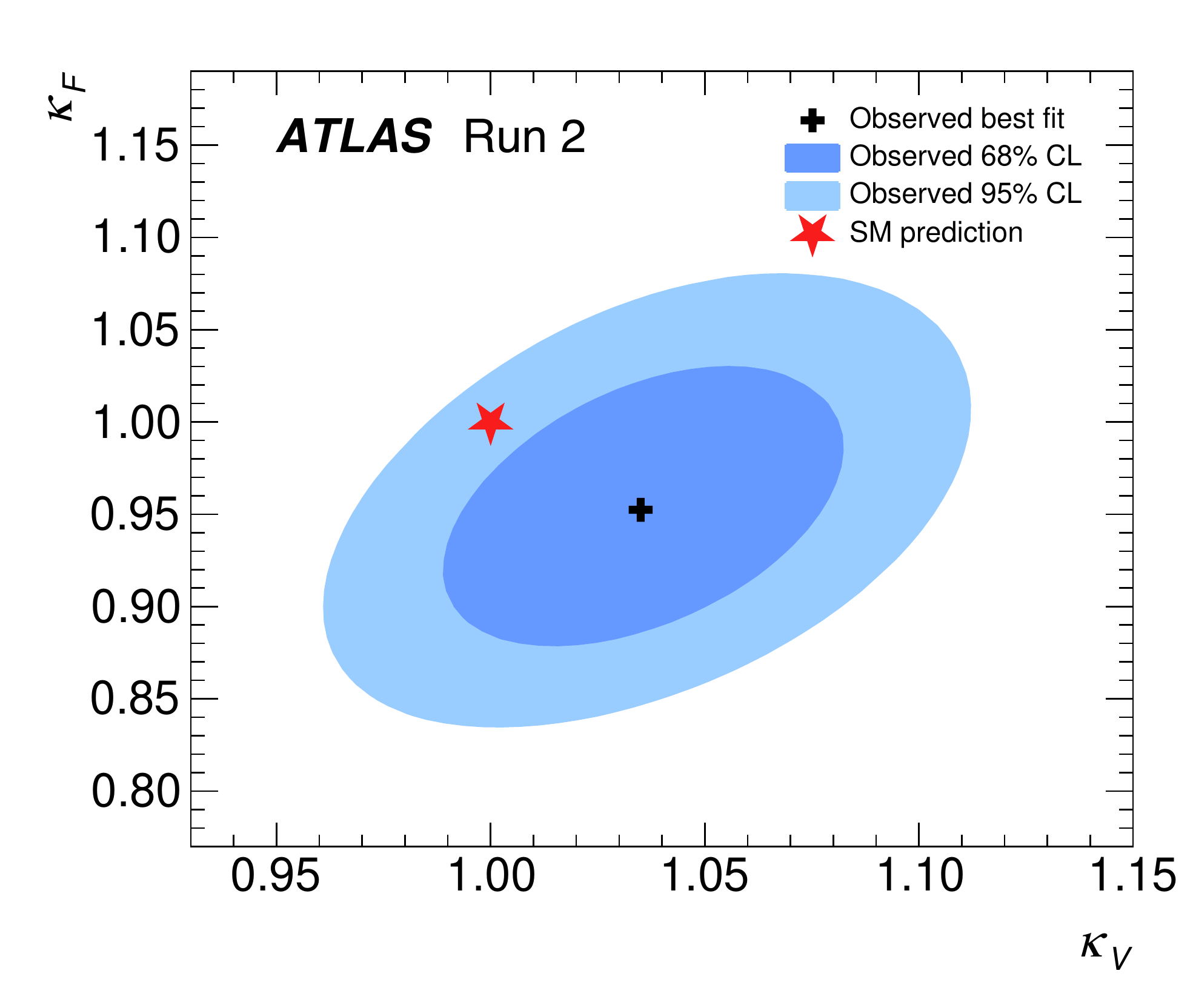}
\caption{\textbf{Negative log-likelihood contours corresponding to  68\% and 95\% CL in the
($\kappa_V$,~$\kappa_F$)~plane.} They are obtained from a combined fit assuming no contributions
from invisible or undetected non-SM Higgs boson decays. The $p$-value for compatibility of
the combined measurement and the SM prediction is 14\%.}
\label{fig:kf_kv}
\end{figure}

In the second class of models, the coupling strength modifiers for $W$, $Z$, $t$, $b$, $c$, $\tau$ and $\mu$ are treated independently. All modifiers are assumed to be positive. It is assumed that only SM particles contribute to the loop-induced processes, and
modifications of the fermion and vector-boson couplings are propagated through the loop
calculations. Invisible or undetected non-SM Higgs boson decays are not considered.
These models allow to test the predicted scaling of the couplings of the Higgs boson to the SM particles as a function of their mass using the reduced coupling strength modifiers $\sqrt{\kappa_V  g_V / 2\text{vev}} =
\sqrt{\kappa_V}  \left(m_V/ \text{vev}\right)$ for weak bosons with a mass $m_V$ and
$\kappa_F  g_F = \kappa_F  m_F / \text{vev}$ for fermions with a mass
$m_F$, where $g_V$ and $g_F$ are the corresponding absolute coupling strengths and
`$\text{vev}$' is the vacuum expectation value of the Higgs field.  Figure~\ref{fig:kappamass} shows the results for
two scenarios: one with the coupling to $c$-quarks
constrained by $\kappa_c = \kappa_t$ in order to cope with the low sensitivity to this
coupling; and the other with $\kappa_c$ left as a free parameter in the fit.  All
measured coupling strength modifiers are found to be compatible with their SM prediction.
When the coupling strength modifier $\kappa_c$ is left unconstrained in the fit, an upper
limit of $\kappa_c < $ 5.7 (7.6) times the SM prediction is observed (expected) at 95\% CL
and the uncertainty in each of the other parameters increases because of the resulting
weaker constraint on the total decay width. This improves the current observed (expected)
limit of $\kappa_c < 8.5$ (12.4) at 95\% CL from the individual measurement of $H\to
\ccbar$ decays~\cite{HIGG-2021-12} despite the relaxed assumptions on other coupling strength modifiers, through constraints coming from the parameterisation of the total Higgs boson decay width that impacts all measurements.
\begin{figure}[!htbp]
\centering
\includegraphics[width=.45\textheight]{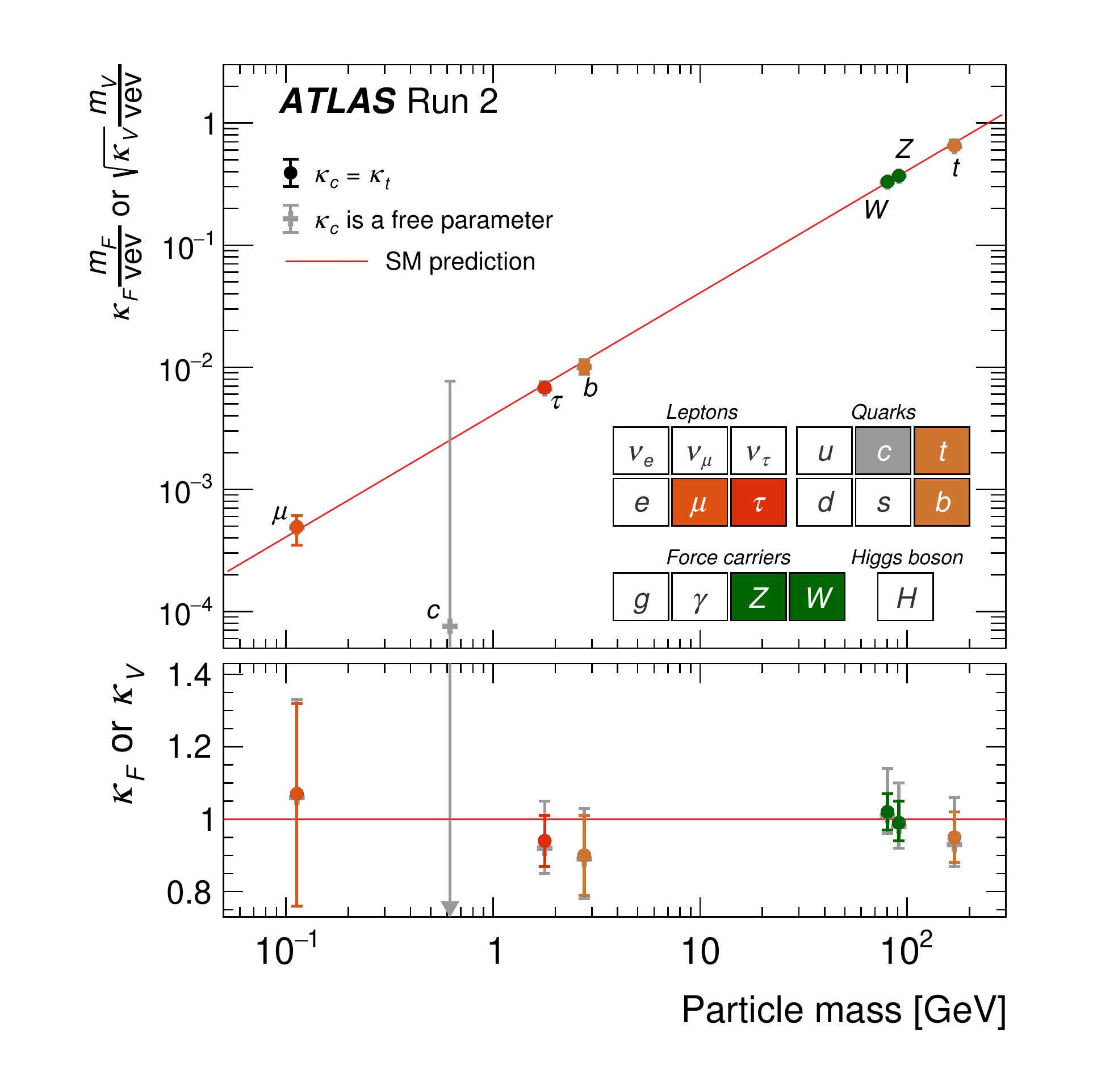}
\caption{
\textbf{Reduced Higgs boson coupling strength modifiers and their uncertainties.} They are defined as $\kappa_F m_F / \text{vev}$ for fermions
($F=t,b,\tau,\mu$) and $\sqrt{\kappa_V} m_V/\text{vev}$ for vector bosons as a
function of their masses $m_F$ and $m_V$. Two fit scenarios with $\kappa_c =
\kappa_t$ (coloured circle markers), or $\kappa_c$ left free-floating in the fit (grey
cross markers) are shown. Loop-induced processes are assumed to have the SM structure, and
Higgs boson decays to non-SM particles are not allowed. The vertical bar on each point
denotes the 68\% confidence interval. The $p$-values for compatibility of the combined measurement and the SM prediction are 56\% and 65\% for the respective scenarios. The lower panel shows the values of the coupling strength modifiers. The grey arrow points in the direction of the best-fit value and the corresponding grey uncertainty bar extends beyond the lower panel range.}
\label{fig:kappamass}
\end{figure}

The third class of models in the $\kappa$-framework closely follows the previous one, but allows for
the presence of non-SM particles in the loop-induced processes. These processes are parameterized
by the effective coupling strength modifiers $\kappa_g$, $\kappa_\gamma$ and
$\kappa_{Z\gamma}$ instead of propagating modifications of the SM particle couplings
through the loop calculations. It is also assumed that any potential effect beyond the SM
does not substantially affect the kinematic properties of the Higgs boson decay products.
The fit results for the scenario in which invisible or undetected
non-SM Higgs boson decays are assumed not to contribute to the total Higgs decay width,
i.e.~$B_{\mathrm{inv.}} = B_{\text{u.}} = 0$, are shown in
Figure~\ref{fig:kappa_generic_bsm} together with the results for the scenario allowing
such decays. To avoid degenerate solutions, the latter constrains $B_{\text{u.}}\geq 0$ and
imposes the additional constraint $\kappa_V \leq 1$ that naturally arises in a variety of
scenarios of physics beyond the SM~\cite{Duhrssen:2004cv, Dobrescu:2012td}. All measured coupling
strength modifiers are compatible with their SM predictions.
\begin{figure}[!htbp]
\centering
\includegraphics[width=.42\textheight]{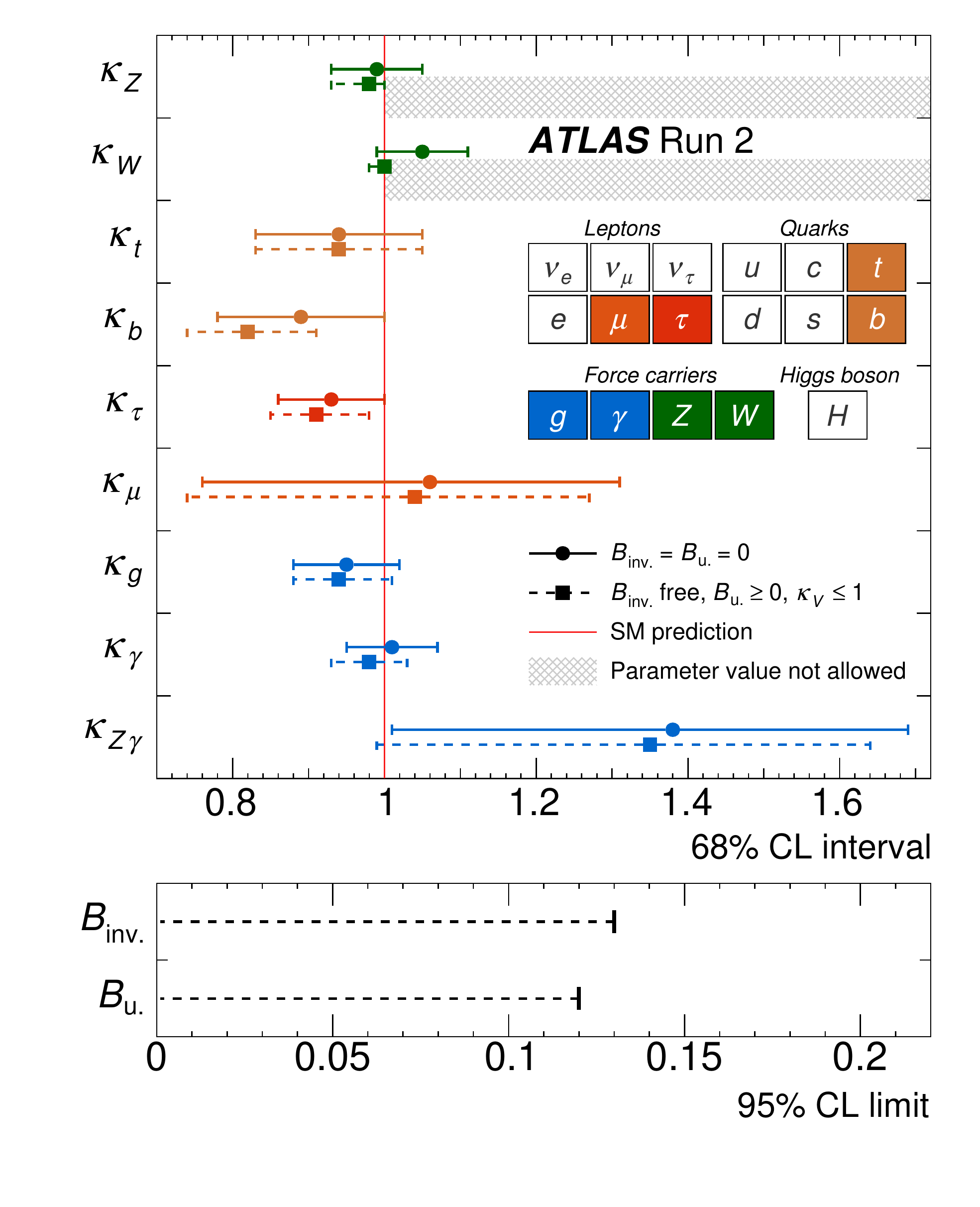}
\caption{\textbf{Reduced coupling strength modifiers and their uncertainties per
particle type with effective photon, $Z\gamma$ and gluon couplings.} The horizontal bars on each point denote the 68\% confidence interval.
The scenario where $B_{\mathrm{inv.}} = B_{\text{u.}} = 0$ is assumed is shown as
solid lines with circle markers. The $p$-value for
compatibility with the SM prediction is 61\% in this case. The scenario
where $B_{\mathrm{inv.}}$ and $B_{\text{u.}}$ are allowed to contribute to the total Higgs boson decay width while
assuming that $\kappa_V\leq 1$ and $B_{\text{u.}}\geq 0$ is shown as dashed lines with
square markers. The lower panel shows the 95\%
CL upper limits on $B_{\mathrm{inv.}}$ and $B_{\text{u.}}$.
}
\label{fig:kappa_generic_bsm}
\end{figure}
When allowing invisible or undetected non-SM Higgs boson decays to contribute to
the total Higgs boson decay width, the previously measured coupling strength modifiers do not
change significantly, while upper limits of $B_{\text{u.}}< 0.12$ (expected 0.21) and
$B_{\mathrm{inv.}} < 0.13$ (expected 0.08) are set at 95\% CL on the
corresponding branching fraction. The latter improves on the current best limit of
$B_{\mathrm{inv.}} <0.145$ (expected 0.103) from direct ATLAS searches~\cite{EXOT-2020-11}.
 
In all tested scenarios, the statistical and the systematic uncertainty contribute almost
equally to the total uncertainty in most of the $\kappa$ parameter measurements. The
exceptions are the $\kappa_\mu$, $\kappa_{Z\gamma}$, $\kappa_c$ and $B_{\text{u.}}$ measurements for which the statistical uncertainty still dominates.

% End of text imported from the .//sections/05_results_kappa.tex input file

% The next lines are included from the .//sections/05_results_stxs.tex input file
Kinematic properties of Higgs boson production probing the internal structure of its couplings are studied in the framework of
simplified template cross sections~\cite{YR4, LesHouches, Berger:2019wnu,
Amoroso:2020lgh}. The framework partitions the phase space of SM Higgs boson production
processes into a set of regions defined by the specific kinematic properties of the Higgs
boson and, where relevant, of the associated jets, $W$ bosons, or $Z$ bosons, as described in Methods.
The regions are defined
so as to provide experimental sensitivity to deviations from the SM predictions, to avoid
large theory uncertainties in these predictions, and to minimize the model-dependence of their extrapolations to the experimentally accessible signal regions.  Signal cross sections
measured in each of the introduced kinematic regions are compared with those predicted when assuming that the branching fractions and kinematic properties of the Higgs boson decay are described by the SM.
 
The results of the simultaneous measurement in 36 kinematic regions are presented in Figure~\ref{fig:stxs_results_norm}.
Compared to previous results with a smaller dataset~\cite{HIGG-2018-57} a much larger number of regions are probed, particularly at high Higgs boson transverse momenta where in many cases the sensitivity to new phenomena beyond the SM is expected to be enhanced. All measurements are consistent with the SM predictions.
\begin{figure}[btp!]
\centering
\includegraphics[width=1.0\textwidth]{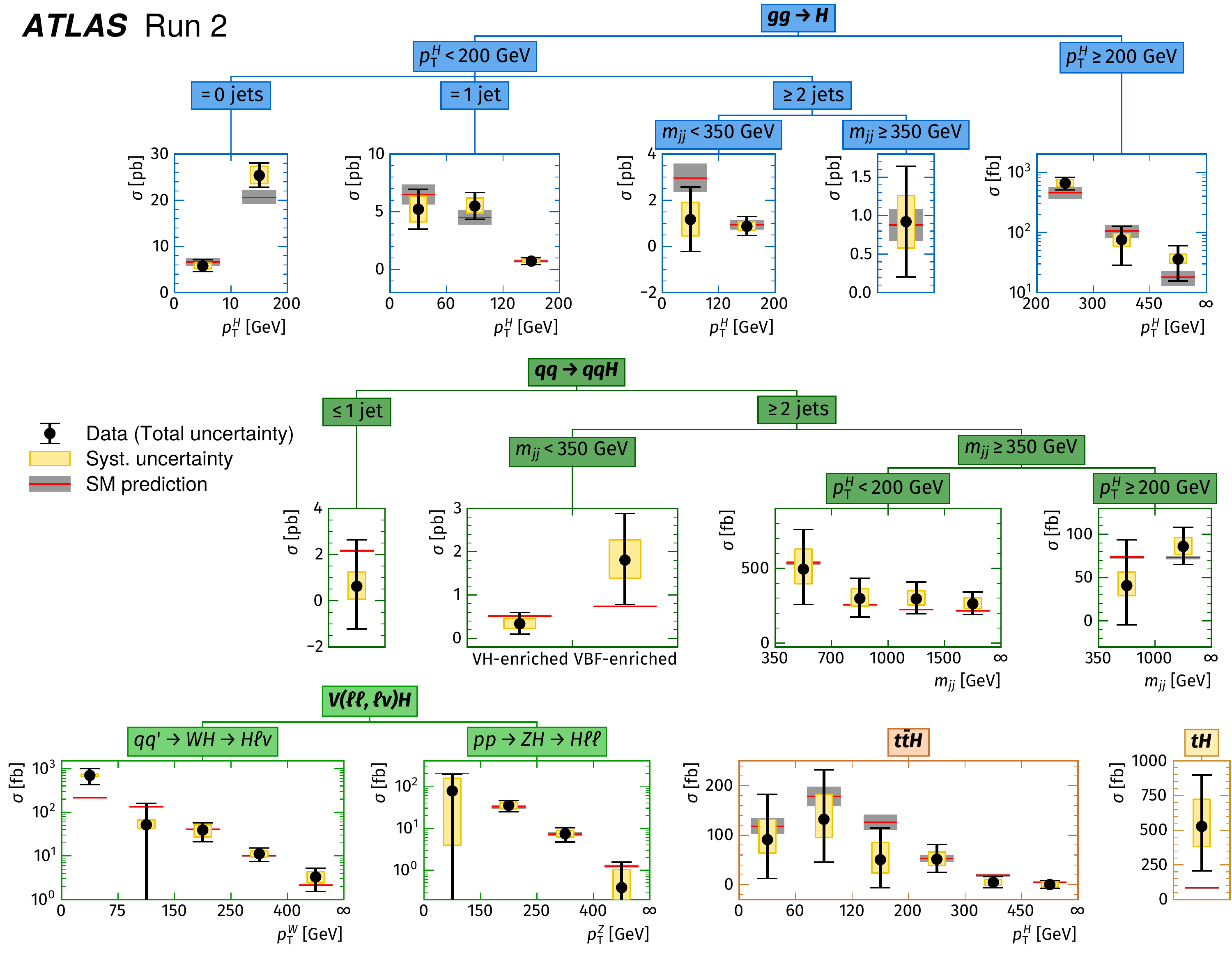}
\caption{\textbf{Observed and predicted Higgs boson production cross sections in different
kinematic regions.} The vertical bar on each point denotes the 68\% confidence interval.  The $p$-value for compatibility of
the combined measurement and the SM prediction is 94\%.
Kinematic regions are defined separately for each production process, based on the jet multiplicity, the transverse momentum of the Higgs ($p_{\textrm{T}}^H$) and vector bosons ($p_{\textrm{T}}^W$ and $p_{\textrm{T}}^Z$) and the two-jet invariant mass ($m_{jj}$).
The `VH-enriched' and `VBF-enriched' regions with the respective requirements of
$m_{jj}\in[60, 120)$ \GeV\ and $m_{jj}\notin[60,120)$ \GeV\ are enhanced in signal events
from $VH$ and VBF productions, respectively.
}
\label{fig:stxs_results_norm}
\end{figure}

% End of text imported from the .//sections/05_results_stxs.tex input file

% End of text imported from the .//sections/05_results.tex input file

\FloatBarrier
\vspace{-15pt}
\section*{Conclusion}
\vspace{-5pt}

% The next lines are included from the .//sections/06_conclusion.tex input file
In summary, the production and decay rates of the Higgs boson were measured using the
dataset collected by the ATLAS experiment during Run~2 of the Large Hadron Collider from 2015 to 2018.  The
measurement results were found to be in excellent agreement with the predictions of the
Standard Model.  In different scenarios, the couplings to the three heaviest fermions, the top quark, the $b$-quark
and the $\tau$-lepton, were measured with uncertainties ranging from about 7\% to 12\% and the couplings to the
weak bosons ($Z$ and $W$) were measured with uncertainties of about 5\%.  In addition, indications are emerging of the presence
of very rare Higgs boson decays into second-generation fermions and into a $Z$ boson and a
photon.  Finally, a comprehensive study of Higgs boson production kinematics was performed
and the results were also found to be compatible with Standard Model predictions.
In the ten years since its discovery, the Higgs boson has undergone many experimental tests that have demonstrated that, so far, its nature is remarkably consistent with the predictions of the Standard Model.
However, some of its key properties
such as the coupling of the Higgs boson to itself remain to be measured.  In addition, some of its rare decay modes have not yet been observed and there is ample room for new phenomena beyond the Standard Model to be discovered.
Substantial progress on these fronts is expected in the future given that detector
upgrades are planned for the coming years, that systematic uncertainties are expected to be
reduced considerably~\cite{Dainese:2703572}, and that the size of the LHC’s dataset is projected to increase by
a factor of 20.

% End of text imported from the .//sections/06_conclusion.tex input file

\section*{Online Content}
\vspace{-5pt}
Any methods, additional references, supplementary information, acknowledgements, and statements of data availability are available in the online paper.

\FloatBarrier

\clearpage
\appendix
\part*{Methods}
\addcontentsline{toc}{part}{Methods}

% The next lines are included from the .//sections/07_methods.tex input file

\section*{Experimental setup}

The ATLAS detector~\cite{PERF-2007-01} consists of an inner tracking detector surrounded by a thin superconducting solenoid, electromagnetic and hadron calorimeters, and a muon spectrometer incorporating three large superconducting air-core toroidal magnets.
 
ATLAS uses a right-handed coordinate system with its origin at the nominal interaction point (IP)
in the center of the detector and the $z$-axis along the beam pipe.
The $x$-axis points from the IP to the center of the LHC ring,
and the $y$-axis points upwards.
Cylindrical coordinates $(r,\phi)$ are used in the transverse plane,
$\phi$ being the azimuthal angle around the $z$-axis.
The pseudorapidity is defined in terms of the polar angle $\theta$ as $\eta = -\ln \tan(\theta/2)$.
 
The inner-detector system (ID) is immersed in a \SI{2}{\tesla} axial magnetic field
and provides charged-particle tracking in the range \(|\eta| < 2.5\).
The high-granularity silicon pixel detector covers the vertex region and typically provides four measurements per track,
the first hit normally being in the insertable B-layer (IBL) installed before Run~2~\cite{ATLAS-TDR-19,PIX-2018-001}.
It is followed by the silicon microstrip tracker (SCT), which usually provides eight measurements per track.
These silicon detectors are complemented by the transition radiation tracker (TRT),
which enables radially extended track reconstruction up to \(|\eta| = 2.0\).
The TRT also provides electron identification information
based on the fraction of hits (typically 30 in total) above a higher energy-deposit threshold corresponding to transition radiation.
 
The calorimeter system covers the pseudorapidity range \(|\eta| < 4.9\).
Within the region \(|\eta|< 3.2\), electromagnetic calorimetry is provided by barrel and
endcap high-granularity lead/liquid-argon (LAr) calorimeters,
with an additional thin LAr presampler covering \(|\eta| < 1.8\)
to correct for energy loss in material upstream of the calorimeters.
Hadron calorimetry is provided by the steel/scintillator-tile calorimeter,
segmented into three barrel structures within \(|\eta| < 1.7\), and two copper/LAr hadron endcap calorimeters.
The solid angle coverage is completed with forward copper/LAr and tungsten/LAr calorimeter modules
optimised for electromagnetic and hadronic energy measurements respectively.
 
The muon spectrometer (MS) comprises separate trigger and
high-precision tracking chambers measuring the deflection of muons in a magnetic field generated by the superconducting air-core toroidal magnets.
The field integral of the toroids ranges between \num{2.0} and \SI{6.0}{\tesla\metre}
across most of the detector.
Three layers of precision chambers, each consisting of layers of monitored drift tubes, covers the region \(|\eta| < 2.7\),
complemented by cathode-strip chambers in the forward region, where the background is highest.
The muon trigger system covers the range \(|\eta| < 2.4\) with resistive-plate chambers in the barrel, and thin-gap chambers in the endcap regions.
 
The performance of the vertex and track reconstruction in the inner detector, the calorimeter resolution in electromagnetic and hadronic calorimeters and the muon momentum resolution provided by the muon spectrometer are given in Ref.~\cite{PERF-2007-01}.
 
Interesting events are selected by the first-level trigger system implemented in custom hardware,
followed by selections made by algorithms implemented in software in the high-level trigger~\cite{TRIG-2016-01}.
The first-level trigger accepts events from the \SI{40}{\MHz} bunch crossings at a rate below \SI{100}{\kHz},
which the high-level trigger further reduces in order to record events to disk at about \SI{1}{\kHz}.

\section*{Statistical framework}
 
The results of the combination presented in this paper are obtained from a likelihood
function defined as the product of the likelihoods of each input measurement. The observed
yield in each category of reconstructed events follows a Poisson distribution whose
parameter is the sum of the expected signal and background contributions. The number of
signal events in any category $k$ is split into the different production and decay
modes:
 
\begin{equation*}
n_k^{\mathrm{signal}} = \mathcal{L}_k\sum_i
\sum_f(\sigma_i B_f)(A\epsilon)_{if}^k,
\label{eq:yields}
\end{equation*}
where the sum indexed by $i$ runs either over the production processes (ggF, VBF, $WH$, $ZH$,
$\ttbar H$, $tH$) or over the set of the measured production kinematic regions, and the sum indexed
by $f$ runs over the decay final states ($ZZ$, $WW$, $\gamma\gamma$, $Z\gamma$, $b\bar{b}$, $c\bar{c}$, $\tau^+\tau^-$, $\mu^+\mu^-$).
The quantity $\mathcal{L}_k$ is the integrated luminosity of the dataset used in category $k$, and
$(A\epsilon)_{if}^k$ is the acceptance times selection efficiency factor for
production process $i$ and decay mode $f$ in category $k$.
Acceptances and efficiencies are obtained from the simulation (corrected by calibration
measurements in control data for the efficiencies). Their values are subject to variations
due to experimental and theoretical systematic uncertainties.
The cross sections $\sigma_i$ and
branching fractions $B_f$ are the parameters of interest of the model. Depending on the model being tested, they are either free
parameters, set to their SM prediction or parameterized as functions of other parameters. All cross sections are defined in the Higgs boson
rapidity range $|y_H| < 2.5$, which is related to the polar angle of the Higgs boson's momentum in the
detector and corresponds approximately to the region of experimental sensitivity.
 
The impact of experimental and theoretical systematic uncertainties on the predicted signal and
background yields is taken into account by nuisance parameters included in the likelihood function.
The predicted signal yields from each production process, the branching fractions and the
signal acceptance in each analysis category are affected by theory uncertainties.
The combined likelihood function is therefore expressed as:
 
\begin{equation*}
L\left(\vec{\alpha}, \vec{\theta}, \text{data}\right) = \prod_{k\in\text{cat}}
\prod_{b\in\text{bins}}   P(n_{k,b} | n^{\text{signal}}_{k,b}(\vec{\alpha}, \vec{\theta}) +
n^{\text{bkg}}_{k,b}(\vec{\theta})) \prod_{\theta\in\vec{\theta}} G(\theta),
\end{equation*}
 
where $n_{k,b}$, $n^{\text{signal}}_{k,b}$ and $n^{\text{bkg}}_{k,b}$
stand for the number of observed events, the number of expected signal events and the
number of expected background events in bin $b$ of analysis category $k$, respectively.
The parameters of interest are noted $\vec{\alpha}$, the nuisance parameters are $\vec{\theta}$, $P$ represents the Poisson distribution, and $G$ stands for Gaussian
constraint terms assigned to the nuisance parameters. Some nuisance parameters are meant to be determined by data alone and do not have any associated constraint term. This is for instance the case of background normalization factors that are fitted in control categories. The effects of nuisance parameters affecting the normalizations of signal and backgrounds in a given category are generally implemented using the multiplicative expression:
\begin{equation*}
n(\theta) = n^0 (1+\sigma)^\theta,
\end{equation*}
where $n^0$ is the nominal expected yield of either signal or background and $\sigma$ the value of the uncertainty. This ensures
that $n(\theta)>0$ even for negative values of $\theta$. For the majority of nuisance
parameters, including all those affecting the shapes of the distributions, a linear
expression is used instead on each bin of the distributions:
\begin{equation*}
n(\theta) = n^0 (1+\sigma \theta).
\end{equation*}
 
The systematic uncertainties are broken down into independent underlying sources, so that when a source affects multiple or all analyses the associated nuisance parameter can be fully correlated across the terms in the likelihood corresponding to these analyses by using common nuisance parameters. This is the case of
systematic uncertainties in the luminosity measurement~\cite{DAPR-2021-01}, in the
reconstruction and selection efficiencies~\cite{PERF-2017-01, TRIG-2018-05, MUON-2018-03,
ATLAS-CONF-2017-029, FTAG-2018-01, FTAG-2020-08, TRIG-2018-08} and in the calibrations of
the energy measurements~\cite{EGAM-2018-01, JETM-2018-05, JETM-2018-06, PERF-2016-07}.
Their effects are propagated coherently by using common
nuisance parameters whenever applicable. Only a few components of the systematic
uncertainties are correlated between the analyses performed using the full Run~2 data and
those using only the 2015 and 2016 data, owing to differences in their assessment, in the
reconstruction algorithms and in software releases. Systematic uncertainties associated
with the modelling of background processes, as well as uncertainties due to the
limited number of simulated events used to estimate the expected signal and background
yields, are treated as being uncorrelated between analyses.
 
Uncertainties in the parton distribution functions are implemented coherently in all input
measurements and all analysis categories~\cite{Butterworth:2015oua}. Uncertainties in
modelling the parton showering into jets of particles affect the signal acceptances and
efficiencies, and are common to all input measurements within a given production
process. Similarly, uncertainties due to missing higher-order quantum chromodynamics (QCD)
corrections are common to a given production process.
Their implementation in the kinematic regions of the simplified template cross-sections
framework results in a total of 66 uncertainty sources, where
overall acceptance effects are separated from migrations between the various bins (e.g.\
between jet multiplicity regions or between dijet invariant mass
regions)~\cite{ATL-PHYS-PUB-2018-035}.
Both the acceptance and signal yield uncertainties affect the signal strength modifier and
coupling strength modifier results, which rely on comparisons of measured and expected yields.
Only acceptance uncertainties affect the cross section and branching fraction
results.
The uncertainties in the Higgs boson branching fractions due to dependencies on SM
parameter values (such as $b$- and $c$-quark masses) and missing higher-order effects are
implemented using the correlation model described in Ref.~\cite{YR4}.
 
In total, over 2600 sources of systematic uncertainty are included in the combined
likelihood. For most of the presented measurements, the systematic uncertainty is expected to be of
similar size or somewhat smaller than the corresponding statistical uncertainty. The
systematic uncertainties are dominant for the parameters which are measured the most
precisely, i.e the global signal strength and the production cross-sections for the ggF
and VBF processes. The expected systematic uncertainty of the global signal strength
measurement (about 5\%) is larger than the statistical uncertainty (3\%), with similar
contributions from the theory uncertainties in signal (4\%) and background modelling
(1.7\%), and from the experimental systematic uncertainty (3\%). The latter is
predominantly composed of the uncertainty in the luminosity measurement (1.7\%), followed
by the uncertainties in electron, jet and $b$-jet reconstruction, data-driven background
modelling, as well as from the limited amount of simulated events (about 1\% each). All
other sources of experimental uncertainty combined contribute an additional 1\%. The
systematic uncertainty in the production cross section of the ggF process is dominated by
experimental uncertainties (3.5\%) followed by signal theory uncertainties (3\%),
compared to a statistical uncertainty of 4\%. For the VBF process, where the statistical
uncertainty is 8\%, the experimental uncertainties are estimated to be 5\% while the
signal theory uncertainties add up to 7\%. Systematic uncertainties are also dominant over
the statistical uncertainties in the measurements of the branching fractions into $W$ pairs and
$\tau$-lepton pairs.
 
Measurements of the parameters of interest use a statistical test based
on the profile likelihood ratio~\cite{Cowan:2010st}:
\begin{equation*}
\Lambda(\vec{\alpha}) = \frac{L(\vec{\alpha},
\vec{\hat{\hat{\theta}}}(\vec{\alpha}))}{L(\vec{\hat{\alpha}},
\vec{\hat{\theta}})},
\end{equation*}
where $\vec{\alpha}$ are the parameters of interest and $\vec{\theta}$
are the nuisance parameters. The $\vec{\hat{\hat{\theta}}}(\vec{\alpha})$ notation
indicates that the nuisance parameters values are the ones that maximize the likelihood
for given values of the parameters of interest. In the denominator, both the parameters of
interest and the nuisance parameters are set to the values ($\vec{\hat{\alpha}}$,
$\vec{\hat{\theta}}$) that unconditionally maximize
the likelihood. The estimates of the parameters $\vec{\alpha}$ are these values
$\vec{\hat{\alpha}}$ that maximize the likelihood ratio.
 
Owing to the usually large number of events selected in the measurements, all results
presented in this paper are obtained in the asymptotic regime where the likelihood
approximately follows a Gaussian distribution.
It was checked in previous iterations of the individual input measurements, for
instance~\cite{HIGG-2011-04}, that this assumption also holds in cases with low event counts by comparing
the results of the asymptotic formulae with those of pseudo-experiments. This confirmed
the results from~\cite{Cowan:2010st} that the Gaussian approximation becomes valid for as few as
$\gtrsim 5$ background events. In the asymptotic regime twice
the negative logarithm
of the profile likelihood $\lambda(\vec{\alpha}) = -2\ln \Lambda(\vec{\alpha})$
follows a $\chi^2$ distribution with a number of degrees of freedom equal to the number of
parameters of interest. Confidence intervals for a given confidence level (CL), usually
68\%, are then defined as the regions fulfilling $\lambda(\vec{\alpha}) <
F_{n}^{-1}(\mathrm{CL})$ where $F_{n}^{-1}$ is the quantile function of the
$\chi^2$ distribution with $n$ degrees of freedom, so $F_{1}^{-1}=1$ (4) for a $1\sigma$
($2\sigma$) CL with one degree of freedom.
The values of the parameters $\vec{\alpha}$ corresponding to these confidence intervals are obtained by scanning the profile likelihood.
Similarly, the $p$-value $p_{\mathrm{SM}} = 1 - F_{n}(\lambda(\vec{\alpha_{\mathrm{SM}}}))$ is used to
test the compatibility of the measurement and the SM prediction. The correlations between
the parameters are estimated by inverting the matrix of the second derivatives of the likelihood.

The \emph{expected} significances and limits are determined using the `Asimov' datasets~\cite{Cowan:2010st}
that are obtained by setting the observed yields to their
expected values when the nuisance parameters are set to the values that maximize the
likelihood $\vec{\hat{\theta}}$.

\section*{Parameterization within the $\kappa$-framework}
 
Within the $\kappa$-framework, the cross section for an individual measurement is parameterized as
\begin{equation*}
\sigma(i  \to H \to f) = \sigma_i \BR_f = \frac{\sigma_i(\vec{\kappa})
\Gamma_f(\vec{\kappa})}{\Gamma_H(\vec{\kappa}, \Binv, \Bund)} ,
\label{eq:kappa:sigma_BR}
\end{equation*}
where $\Gamma_f$ is the partial width for a Higgs boson decay to the final state $f$ and
$\Gamma_H$ is the total decay width of the Higgs boson. The total width is given by the
sum of the partial widths of all the decay modes included. Contributions to the total Higgs boson decay width due to phenomena beyond the SM may manifest themselves as a value of coupling strength modifier $\kappa_{p}$ differing from one, or a value of \Binv\ or \Bund\ differing from zero. The Higgs boson total width is then expressed as $\Gamma_{H}(\vec{\kappa}, \Binv, \Bund) = \kappa_H^2(\vec{\kappa}, \Binv, \Bund) \, \Gamma_{H}^{\text{SM}}$ with
\begin{equation*}
\kappa_H^2(\vec{\kappa}, \Binv, \Bund) = \frac{\sum_p \BR_p^\text{SM} \kappa_p^2}{(1 - \Binv - \Bund)} .
\label{eq:CH:1}
\end{equation*}
Higgs boson production cross sections and partial and total decay widths are parameterized
in terms of the coupling strength modifiers as shown in Table~9 of
Ref.~\cite{HIGG-2018-57}. An improved parameterization including additional sub-leading contributions is used in this paper to
match the increased precision of the measurements.

\section*{Kinematic regions probing Higgs boson production}
 
The definitions of kinematic regions for the precision study of Higgs boson production in the framework
of simplified template cross sections~\cite{YR4,LesHouches,Berger:2019wnu,Amoroso:2020lgh}
are based on the predicted properties of particles generated in a given production
process. The partitioning follows the so-called Stage-1.2 scheme, which features a
slightly finer granularity than the Stage-1.1 scheme of Ref.~\cite{Berger:2019wnu} and
introduces the Higgs boson transverse momentum categories for the $t\bar{t}H$ production
process. Higgs bosons are required to be produced with rapidity $|y_H| < 2.5$. Associated
jets of particles are constructed from all stable particles with a lifetime greater than
$10\,\ps$, excluding the decay products of the Higgs boson and leptons from $W$ and $Z$
boson decays, using the anti-$k_t$ algorithm~\cite{Cacciari:2008gp} with a jet radius parameter $R = 0.4$, and must have a transverse momentum $\pTj\ > 30\,\GeV$. SM predictions are assumed for the kinematic properties of Higgs boson decays. Phenomena
beyond the SM can substantially modify these properties, and thus the acceptance of the signal, especially for the $WW$ or $ZZ$ decay modes, and this should be considered when using these measurements for the relevant interpretations.
 
Higgs boson production is first classified according to the nature of the initial state and the associated particles, the latter including the decay products of $W$ and $Z$ bosons if they are present. These classes are: $t\bar{t}H$ and $tH$ processes; $qq'\to Hqq'$ processes, with contributions from both VBF and quark-initiated $VH$ (where $V=W, Z$) production with a hadronic decay of the vector boson; $pp\to VH$ production with a leptonic decay of the vector boson ($V(\ell\ell,\ell\nu)H$), including $gg\to ZH \to \ell\ell H$ production; and finally the ggF process combined with $gg\to ZH \to q\bar{q}H$ production to form a single $gg\to H$ process. The contribution of the $b\bar{b}H$ production process is taken into account as a $1\%$~\cite{YR4} increase of the \ggtoH\ yield in each kinematic region, since the acceptances for both processes are similar for all input analyses~\cite{YR4}.
 
The input measurements in individual decay modes provide only limited sensitivity to the cross section in some of the regions of the Stage-1.2 scheme, mainly because of the small number of events in some of these regions. In other cases, they only provide sensitivity to a combination of these regions, leading to strongly correlated measurements. To mitigate these effects, some of the Stage-1.2 kinematic regions were merged for the combined measurement.
 
Compared to individual input measurements, systematic theory uncertainties associated with
the signal predictions have been updated for the combination to closely follow the
granularity of the Stage-1.2 scheme. The QCD scale uncertainties in ggF production were
updated for all input channels which are sensitive to this production process. Out of 18 uncertainty sources in total, two account for overall fixed-order and resummation effects, two cover the migrations between different jet multiplicity bins, seven are associated with the modelling of the Higgs boson transverse momentum ($p_\text{T}^H$) in different phase-space regions, four account for the uncertainty in the distribution of the dijet invariant mass ($m_{jj}$) variable, one covers the modelling of the Higgs boson plus two leading jets transverse momentum ($p_\text{T}^{Hjj}$) distribution in the $\ge$2-jet region, one pertains to modelling of the distribution of the Higgs boson plus one jet transverse momentum ($p_\text{T}^{Hj}$) divided by $p_\text{T}^H$ in the high-$p_\text{T}^H$ region, and finally, the last takes into account the uncertainty from the choice of top-quark mass scheme. Theory uncertainties for the $qq'\to Hqq'$ and $t\bar{t}H$ processes are defined as in Ref.~\cite{HIGG-2018-28}, and those of the $V(\ell\ell,\ell\nu)H$ kinematic region follow the scheme described in Ref.~\cite{ATL-PHYS-PUB-2018-035}. For the kinematic regions defined by the merging of several Stage-1.2 regions, the signal acceptance factors are determined assuming that the relative fractions in each Stage-1.2 region are given by their SM values, and the uncertainties predicted by the SM in these fractions are taken into account.
 
\section*{Author contributions}
\vspace{-5pt}
All authors have contributed to the publication, being variously involved in the design and the construction of the detectors, in writing software, calibrating subsystems, operating the detectors and acquiring data, and finally analysing the processed data. The ATLAS Collaboration members discussed and approved the scientific results. The manuscript was prepared by a subgroup of authors appointed by the collaboration and subject to an internal collaboration-wide review process. All authors reviewed and approved the final version of the manuscript.
 
\section*{Competing financial interests}
\vspace{-5pt}
The authors declare no competing financial interests.
 
\section*{Data availability}
\vspace{-5pt}
The experimental data that support the findings of this study are available in HEPData with the identifier \href{https://doi.org/10.17182/hepdata.130266}{https://doi.org/10.17182/hepdata.130266}.
 
\section*{Code availability}
\vspace{-5pt}
The ATLAS data reduction software is available at the following link: \href{https://zenodo.org/record/4772550}{https://zenodo.org/record/4772550}.  Statistical modeling and analysis is based on the ROOT software and its embedded RooFit and RooStats modules, available at the following link:  \href{https://zenodo.org/record/3895852}{https://zenodo.org/record/4772550}.  Code to configure these statistical tools and to process their output is available upon request.

\FloatBarrier

% End of text imported from the .//sections/07_methods.tex input file

\newpage
\section*{Acknowledgements}

% The next lines are included from the .//acknowledgements/Acknowledgements.tex input file

We thank CERN for the very successful operation of the LHC, as well as the
support staff from our institutions without whom ATLAS could not be
operated efficiently.
 
We acknowledge the support of
ANPCyT, Argentina;
YerPhI, Armenia;
ARC, Australia;
BMWFW and FWF, Austria;
ANAS, Azerbaijan;
CNPq and FAPESP, Brazil;
NSERC, NRC and CFI, Canada;
CERN;
ANID, Chile;
CAS, MOST and NSFC, China;
Minciencias, Colombia;
MEYS CR, Czech Republic;
DNRF and DNSRC, Denmark;
IN2P3-CNRS and CEA-DRF/IRFU, France;
SRNSFG, Georgia;
BMBF, HGF and MPG, Germany;
GSRI, Greece;
RGC and Hong Kong SAR, China;
ISF and Benoziyo Center, Israel;
INFN, Italy;
MEXT and JSPS, Japan;
CNRST, Morocco;
NWO, Netherlands;
RCN, Norway;
MEiN, Poland;
FCT, Portugal;
MNE/IFA, Romania;
MESTD, Serbia;
MSSR, Slovakia;
ARRS and MIZ\v{S}, Slovenia;
DSI/NRF, South Africa;
MICINN, Spain;
SRC and Wallenberg Foundation, Sweden;
SERI, SNSF and Cantons of Bern and Geneva, Switzerland;
MOST, Taiwan;
TENMAK, T\"urkiye;
STFC, United Kingdom;
DOE and NSF, United States of America.
In addition, individual groups and members have received support from
BCKDF, CANARIE, Compute Canada and CRC, Canada;
PRIMUS 21/SCI/017 and UNCE SCI/013, Czech Republic;
COST, ERC, ERDF, Horizon 2020 and Marie Sk{\l}odowska-Curie Actions, European Union;
Investissements d'Avenir Labex, Investissements d'Avenir Idex and ANR, France;
DFG and AvH Foundation, Germany;
Herakleitos, Thales and Aristeia programmes co-financed by EU-ESF and the Greek NSRF, Greece;
BSF-NSF and MINERVA, Israel;
Norwegian Financial Mechanism 2014-2021, Norway;
NCN and NAWA, Poland;
La Caixa Banking Foundation, CERCA Programme Generalitat de Catalunya and PROMETEO and GenT Programmes Generalitat Valenciana, Spain;
G\"{o}ran Gustafssons Stiftelse, Sweden;
The Royal Society and Leverhulme Trust, United Kingdom.
 
The crucial computing support from all WLCG partners is acknowledged gratefully, in particular from CERN, the ATLAS Tier-1 facilities at TRIUMF (Canada), NDGF (Denmark, Norway, Sweden), CC-IN2P3 (France), KIT/GridKA (Germany), INFN-CNAF (Italy), NL-T1 (Netherlands), PIC (Spain), ASGC (Taiwan), RAL (UK) and BNL (USA), the Tier-2 facilities worldwide and large non-WLCG resource providers. Major contributors of computing resources are listed in Ref.~\cite{ATL-SOFT-PUB-2023-001}.

% End of text imported from the .//acknowledgements/Acknowledgements.tex input file

\clearpage
 
\part*{Extended Data}
\addcontentsline{toc}{part}{Extended Data}

% The next lines are included from the .//sections/08_extendeddata.tex input file

\begin{table}[!htbp]
\caption{\textbf{Summary of the input measurements employed for the combined fit
results.}
Listed are the measured decay modes, targeted production processes and integrated luminosity ($\mathcal{L}$) used for each input analysis of the combination.
The references for the input measurements and information about which combined
measurements they enter are also provided. The $\ttH+\tH$ measurement in the multilepton
final state~\cite{HIGG-2017-02} is complementary to other $\ttH+\tH$ measurements in the $WW$, $ZZ$ and $\tau\tau$ decay modes, with no overlapping events selected by these measurements.
}
\centering

{  \renewcommand{\arraystretch}{1.0}
\begin{tabular}{lrlcl}
\toprule
Decay mode            & Targeted production processes & $\mathcal{L}$ [fb$^{-1}$] &
Ref. & Fits deployed in\\
\midrule
\Hyy\                       & $\ggF, \VBF, \WH, \ZH, \ttH, tH$ & $139$ &
\cite{HIGG-2020-16} & All\\[2ex]
 
\hzz                       & $\ggF, \VBF, \WH+\ZH, \ttH+tH$ & 139 & \cite{HIGG-2018-28} & All \\
& $\ttH+\tH$  (multilepton)
& ~~36.1 & \cite{HIGG-2017-02}
&  All but fit of
kinematics\\[2ex]
 
\hww                     & $\ggF,\VBF$                  & 139 &
\cite{HIGG-2021-20-bib} & All\\
& $\WH, \ZH$                  & ~~36.1 & \cite{HIGG-2017-14} & All but fit of kinematics\\
& $\ttH+\tH$    (multilepton)
& ~~36.1 & \cite{HIGG-2017-02}
&  All but fit of
kinematics\\[2ex]
 
$H\to Z\gamma$                        & inclusive & $139$ & \cite{HIGG-2018-42} &
All but fit of kinematics\\[2ex]

\hbb    & $\WH, \ZH$               & 139 & \cite{HIGG-2018-51, HIGG-2018-52} & All\\
& $\VBF$                   & 126 & \cite{HIGG-2019-04} & All \\
& $\ttH+\tH$                   & 139 & \cite{HIGG-2020-23} & All \\
& inclusive                     & 139 &
\cite{HIGG-2021-08}& Only for fit of kinematics\\[2ex]
 
\htt              & $\ggF,\VBF,\WH+\ZH,\ttH+\tH$    & 139 & \cite{HIGG-2019-09} & All \\
& $\ttH+\tH$ (multilepton)
& ~~36.1 & \cite{HIGG-2017-02} &  All
but fit of kinematics\\[2ex]
 
\hmm                               & $\ggF+\ttH+\tH,\VBF+\WH+\ZH$     & 139 &
\cite{HIGG-2019-14} & All but fit of kinematics\\[2ex]
 
$H\to \ccbar$                         & $\WH+\ZH$     & 139 & \cite{HIGG-2021-12} &
Only for free-floating $\kappa_c$\\[2ex]

$H\to \text{invisible}$                       & $\VBF$                   &
139 & \cite{EXOT-2020-11} & $\kappa$ models with $B_{\text{u.}} \&~B_{\text{inv.}}$\\
& $\ZH$
& 139 &
\cite{HIGG-2018-26} &  $\kappa$ models with $B_{\text{u.}} \&~B_{\text{inv.}}$\\

\bottomrule
\end{tabular}
}
\label{tab:inputs}
\end{table}

\FloatBarrier

% End of text imported from the .//sections/08_extendeddata.tex input file

\clearpage
\printbibliography
 
\clearpage
\input{atlas_authlist}

\end{document}

%% file: atlas_authlist.tex
% ATLAS Collaboration author list
% Reference date of HIGG-2021-23 is 2022-03-01
% Author list last updated on date 13-OCT-23
% Data extracted on 13-Oct-2023 for paper reference HIGG-2021-23
% at 8:05pm
 
\begin{flushleft}
\hypersetup{urlcolor=black}
{\Large The ATLAS Collaboration}

\bigskip

\AtlasOrcid[0000-0002-6665-4934]{G.~Aad}$^\textrm{\scriptsize 101}$,
\AtlasOrcid[0000-0002-5888-2734]{B.~Abbott}$^\textrm{\scriptsize 119}$,
\AtlasOrcid[0000-0002-7248-3203]{D.C.~Abbott}$^\textrm{\scriptsize 102}$,
\AtlasOrcid[0000-0002-1002-1652]{K.~Abeling}$^\textrm{\scriptsize 55}$,
\AtlasOrcid[0000-0002-8496-9294]{S.H.~Abidi}$^\textrm{\scriptsize 29}$,
\AtlasOrcid[0000-0002-9987-2292]{A.~Aboulhorma}$^\textrm{\scriptsize 35e}$,
\AtlasOrcid[0000-0001-5329-6640]{H.~Abramowicz}$^\textrm{\scriptsize 150}$,
\AtlasOrcid[0000-0002-1599-2896]{H.~Abreu}$^\textrm{\scriptsize 149}$,
\AtlasOrcid[0000-0003-0403-3697]{Y.~Abulaiti}$^\textrm{\scriptsize 116}$,
\AtlasOrcid[0000-0003-0762-7204]{A.C.~Abusleme~Hoffman}$^\textrm{\scriptsize 136a}$
% ...
\bigskip
\\

% ...
$^{29}$Physics Department, Brookhaven National Laboratory, Upton NY; United States of America.\\
% ...
$^{35}$$^{(a)}$Facult\'e des Sciences Ain Chock, R\'eseau Universitaire de Physique des Hautes Energies - Universit\'e Hassan II, Casablanca;$^{(b)}$Facult\'{e} des Sciences, Universit\'{e} Ibn-Tofail, K\'{e}nitra;$^{(c)}$Facult\'e des Sciences Semlalia, Universit\'e Cadi Ayyad, LPHEA-Marrakech;$^{(d)}$LPMR, Facult\'e des Sciences, Universit\'e Mohamed Premier, Oujda;$^{(e)}$Facult\'e des sciences, Universit\'e Mohammed V, Rabat;$^{(f)}$Institute of Applied Physics, Mohammed VI Polytechnic University, Ben Guerir; Morocco.\\
% ...
$^{55}$II. Physikalisches Institut, Georg-August-Universit\"{a}t G\"ottingen, G\"ottingen; Germany.\\
% ...
$^{101}$CPPM, Aix-Marseille Universit\'e, CNRS/IN2P3, Marseille; France.\\
$^{102}$Department of Physics, University of Massachusetts, Amherst MA; United States of America.\\
% ...
$^{116}$Department of Physics, New York University, New York NY; United States of America.\\
% ...
$^{119}$Homer L. Dodge Department of Physics and Astronomy, University of Oklahoma, Norman OK; United States of America.\\
% ...
$^{136}$$^{(a)}$Departamento de F\'isica, Pontificia Universidad Cat\'olica de Chile, Santiago;$^{(b)}$Millennium Institute for Subatomic physics at high energy frontier (SAPHIR), Santiago;$^{(c)}$Instituto de Investigaci\'on Multidisciplinario en Ciencia y Tecnolog\'ia, y Departamento de F\'isica, Universidad de La Serena;$^{(d)}$Universidad Andres Bello, Department of Physics, Santiago;$^{(e)}$Instituto de Alta Investigaci\'on, Universidad de Tarapac\'a, Arica;$^{(f)}$Departamento de F\'isica, Universidad T\'ecnica Federico Santa Mar\'ia, Valpara\'iso; Chile.\\
% ...
$^{149}$Department of Physics, Technion, Israel Institute of Technology, Haifa; Israel.\\
$^{150}$Raymond and Beverly Sackler School of Physics and Astronomy, Tel Aviv University, Tel Aviv; Israel.\\
% ...
$^{a}$ Also Affiliated with an institute covered by a cooperation agreement with CERN.\\
$^{b}$ Also at An-Najah National University, Nablus; Palestine.\\
$^{c}$ Also at Borough of Manhattan Community College, City University of New York, New York NY; United States of America.\\
$^{d}$ Also at Bruno Kessler Foundation, Trento; Italy.\\
$^{e}$ Also at Center for High Energy Physics, Peking University; China.\\
$^{f}$ Also at Centro Studi e Ricerche Enrico Fermi; Italy.\\
% ...

\end{flushleft}

% Created with Glance <Atlas.Glance@cern.ch>